\documentclass{article}

\font\scap=cmcsc10 \hfuzz=5cm
\font\scap=cmcsc10

\newcount\eqnumber
\def\neweq{{{(\the\eqnumber)}}\global\advance\eqnumber by 1}
\def\eqdef#1{\eqno\xdef#1{\the\eqnumber}\neweq}
\def\newaeq{{{(\the\eqnumber { a})}}\global\advance\eqnumber by 1}
\def\eqdaf#1{\eqno\xdef#1{\the\eqnumber}\newaeq}
\def\eqdisp#1{\xdef#1{\the\eqnumber}\neweq}
\def\eqdasp#1{\xdef#1{\the\eqnumber}\newaeq}

\newcount\refnumber
\def\newref{{\the\refnumber}\global\advance\refnumber by 1}
\def\refdef#1{{\xdef#1{\the\refnumber}}\newref}

\begin{document}

\centerline{\bf Multiplicative equations related to the affine Weyl group E$_8$}
\bigskip
\bigskip{\scap B. Grammaticos} and {\scap A. Ramani}
\quad{\sl IMNC, Universit\'e Paris VII \& XI, CNRS, UMR 8165, B\^at. 440, 91406 Orsay, France}

\medskip{\scap R. Willox} \quad
{\sl Graduate School of Mathematical Sciences, the University of Tokyo, 3-8-1 Komaba, Meguro-ku, 153-8914 Tokyo, Japan }

\medskip{\scap J. Satsuma}
\quad{\sl  Department of Mathematical Engineering, Musashino University, 3-3-3 Ariake, Koto-ku, 135-8181 Tokyo, Japan}

\bigskip
{\scap Abstract}
\smallskip
We derive integrable equations starting from autonomous mappings with a general form inspired by the multiplicative systems associated to the affine Weyl group E$_8^{(1)}$. Five such systems are obtained, three of which turn out to be linearisable and the remaining two are integrable in terms of elliptic functions. In the case of the linearisable mappings we derive nonautonomous forms which contain a free function of the dependent variable and we present the linearisation in each case. The two remaining systems are deautonomised to new discrete Painlev\'e equations. We show that these equations are in fact special forms of much richer systems associated to the affine Weyl groups E$_7^{(1)}$ and E$_8^{(1)}$ respectively. 



\bigskip
1. {\scap Introduction}
\medskip

Deriving discrete Painlev\'e equations [\refdef\dps] is a task that may appear straightforward, but which can easily lead to prohibitively lengthy calculations. The method that has led to the derivation of the majority of discrete Painlev\'e equations is known as deautonomisation [\refdef\desoto]. This procedure consists in obtaining non-autonomous extensions of integrable mappings, by using some suitable integrability criterion. The two discrete integrability criteria customarily used for this task are singularity confinement [\refdef\sincon] and algebraic entropy [\refdef\hiv]. Both, however, have their advantages and shortcomings. In singularity confinement one must examine the behaviour of {\sl all} singularities: missing a  single one may lead to wrong results. Moreover there does not exist an {\sl a priori} clear indication as to the length of the singularity patterns, i.e. at which iteration one must impose the confinement constraints. Handling these difficulties the easy way requires an essential amount of experience.  The hard way would be to perform the full algebro-geometric analysis [\refdef\algeo] of the system at hand, a process which can, in principle, remove the ambiguities of the simple singularity analysis. The algebraic entropy approach on the other hand is more straightforward: one has to study the degree growth of the iterates of some initial condition and require that it be slow enough. However, the constraints for curbing the exponential growth (which would signal non-integrability)  must be implemented at the appropriate iteration step, something which presents the same difficulty as in the case of singularity confinement. Moreover, these constraints usually affect all the parameters of the equation at the same time, thus leading to expressions that are very difficult to disentangle. The problem of the entanglement of the parameters in the integrability constraints, furnished by the algebraic entropy approach, becomes particularly crucial in the case of discrete Painlev\'e equations which have many parameters like those associated to the affine Weyl group E$_8^{(1)}$. The general setting for the description of these equations was furnished by the groundbreaking work of Sakai [\refdef\sakai]. The detailed construction of the generic equations related to  E$_8^{(1)}$ was given in [\refdef\grand]. 

The calculational difficulties which one encounters when studying discrete Painlev\'e equations associated to E$_8^{(1)}$ have been the main reason for the relative paucity of results on these interesting systems.  The geometry of the eight-parameter discrete Painlev\'e equations has been given in [\grand], while the first construction of an elliptic Painlev\'e equation was presented in [\refdef\ohta]. A first form for the generic elliptic Painlev\'e equation was derived by Murata [\refdef\murata] and it was later somewhat simplified in [\refdef\yoneda]. Recently, however, there has been substantial progress in the study of discrete Painlev\'e equations associated to E$_8^{(1)}$. Noumi and collaborators derived a form for the general elliptic Painlev\'e equation [\refdef\tsuji]. An explicit construction of several E$_8^{(1)}$-related equations of additive type was given in [\refdef\addeight], and in [\refdef\ellip] it was shown how those results could be transposed to the multiplicative and elliptic case.  This recent progress was due to the introduction of new forms which recast the generic E$_8^{(1)}$-associated equation into a more convenient form, more amenable to calculations. 

The first such form is the one we dubbed trihomographic. It has the form
$${x_{n+1}-a_n\over x_{n+1}-b_n}{x_{n-1}-c_n\over x_{n-1}-d_n}{x_{n}-e_n\over x_{n}-f_n}=g_n,\eqdef\eqi$$
and was directly inspired by the basic Miura transformation [\grand] in the E$_8^{(1)}$ space. As was shown in [\addeight] this trihomographic form is perfectly equivalent to the generic additive and multiplicative E$_8^{(1)}$-associated equations, provided one chooses the rational expression on their right-hand side in the appropriate way, taking $g_n=1$. 
In the multiplicative case, which is of interest to us here,  we have the following trihomographic form
$$\displaylines{{x_{n+1}-k_nz_nz_{n-1}-{1\over k_nz_nz_{n-1}}\over x_{n+1}-{z_nz_{n-1}\over k_n}-{k_n\over z_nz_{n-1}}}{x_{n-1}-k_nz_nz_{n+1}-{1\over k_nz_nz_{n+1}}\over x_{n-1}-{z_nz_{n+1}\over k_n}-{k_n\over z_nz_{n+1}}}{x_{n}-{z_n^2z_{n-1}z_{n+1}\over k_n}-{k_n\over z_n^2z_{n-1}z_{n+1}}\over x_{n}-k_nz_n^2z_{n-1}z_{n+1}-{1\over k_nz_n^2z_{n-1}z_{n+1}}}=1,\hfill\cr\hfill
\eqdisp\eqii\cr}$$
which is identical to
$${(x_{n+1}z_{n+1}z_n-x_{n})(x_{n-1}z_{n-1}z_n-x_{n})-(z_{n+1}^2z_n^2-1)(z_{n-1}^2z_n^2-1)\over(x_{n+1}-z_{n+1}z_nx_{n})(x_{n-1}-z_{n-1}z_nx_{n})-(z_{n+1}^2z_n^2-1)(z_{n-1}^2z_n^2-1)/(z_{n+1}z_n^2z_{n-1})}=R(x_n),\eqdef\eqiii$$ 
with
$$R(x_n)={x_n-z_{n+1}z_n^2z_{n-1}(k_n+1/k_n)\over x_nz_{n+1}z_n^2z_{n-1}-k_n-1/k_n},\eqdef\eqiv$$
and where $k_n$ is an (as yet unspecified) function of $n$.
Moreover, as shown in [\refdef\trijmp], the trihomographic representation is not limited to the E$_8^{(1)}$-associated discrete Painlev\'e equations but provides a novel and extremely convenient way to represent all discrete Painlev\'e equations, even the simplest ones. 

The second breakthrough in the study of the E$_8^{(1)}$-associated equations came with the realisation that the equations can be cast in a much simpler form provided one introduces an ancillary variable. The latter is defined through
$$x_n=\xi_n^2,\qquad x_n=\xi_n+{1\over\xi_n}\quad{\rm or}\quad x_n={\theta_1^2(\xi_n)\over\theta_0^2(\xi_n)},\eqdef\eqv$$
for the additive, multiplicative and elliptic equations respectively (the $\theta_i$ being Jacobi theta functions). For instance, in the multiplicative case it is easy to show that the expression
$${x_{n+1}-{\xi_nz_n^{-1}z_{n+1}^{-1}}-{\xi_n^{-1}z_nz_{n+1}}\over x_{n+1}-\xi_nz_nz_{n+1}-{\xi_n^{-1}z_n^{-1}z_{n+1}^{-1}}}\,{x_{n-1}-{\xi_nz_n^{-1}z_{n-1}^{-1}}-{\xi_n^{-1}z_nz_{n-1}}\over x_{n-1}-\xi_nz_nz_{n-1}-{\xi_n^{-1}z_n^{-1}z_{n-1}^{-1}}}={\prod_{i=1}^8(\xi_n-z_n\mu_n^i)\over\prod_{i=1}^8(z_n\mu_n^i\xi_n-1)},\eqdef\eqvi$$
is equivalent to the canonical E$_8^{(1)}$-associated multiplicative Painlev\'e equation, i.e. equation (\eqiii), with right-hand side 
$$\displaylines{R(x_n)= z_{n+1}z_n^2z_{n-1}
{x_n^4-x_n^3Q_1-x_n^2(Q_8-Q_2+3)+x_n(Q_7-Q_3+2Q_1)+Q_8-Q_6+Q_4-Q_2+1\over x_n^4Q_8-x_n^3Q_7-x_n^2(3Q_8-Q_6+1)+x_n(2Q_7-Q_5+Q_1)+Q_8-Q_6+Q_4-Q_2+1}.~\hfill\cr\hfill \eqdisp\eqvii\cr}$$
The $Q_k$ are the elementary symmetric functions constructed out of the quantities $z_n\mu_n^i$, with $\mu^i$ being eight parameters which may depend on the independent variable. This factorised form makes the application of singularity analysis quite straightforward, simplifying greatly the otherwise prohibitively bulky calculations.

In [\refdef\ancil] we presented this new approach to E$_8^{(1)}$-associated discrete Painlev\'e equations starting from the trihomographic representation. This has allowed, among others, to obtain the general form of the elliptic discrete Painlev\'e equation [\refdef\ellip] in a most compact way. The ancillary variable was also introduced by Kajiwara, Noumi and Yamada in their monumental work [\refdef\kanoya] which contains a cornucopia of results on discrete Painlev\'e equations. 

In a recent paper of ours [\refdef\miura] we addressed the question of the derivation of discrete Painlev\'e equations from a different point of view. Namely, we postulated a form inspired by that of additive E$_8^{(1)}$-associated equations but chose a particularly simple right-hand side. Starting from autonomous forms we obtained several integrable candidates which we then proceeded to deautonomise and for which we showed how they could either be related to discrete Painlev\'e equations associated to the affine Weyl groups E$_7^{(1)}$ and E$_6^{(1)}$, or actually linearised if a study of their growth made them candidates for linearisability.
In this paper we shall adopt a similar strategy albeit tailored to equations of multiplicative type, which as we shall see, is not an entirely trivial matter. While in general the study of additive and multiplicative equations proceeds in parallel, with the results obtained for one type being easily transcribed to the other one, this is not the case here, given the form of the right-hand side we postulate. We shall show that our approach leads to new integrable systems, both discrete Painlev\'e equations and linearisable mappings. 

\bigskip
2. {\scap A quest for integrable mappings}
\medskip

In order to investigate the possible existence of new integrable mappings of multiplicative type we start from an ansatz inspired by the form (\eqiii). Moreover, as is customary in the deautonomisation approach, we start from an autonomous form. Choosing a very simple right-hand side our starting point is 
$${(x_{n+1}z^2-x_{n})(x_{n-1}z^2-x_{n})-(z^4-1)^2\over(x_{n+1}-z^2x_{n})(x_{n-1}-z^2x_{n})-(z^2-1/z^2)^2}=fz^N,\eqdef\eqviii$$ 
which we investigate from the point of view of integrability using the algebraic entropy method. We start from initial conditions $x_0$ and $x_1=p/q$ and we calculate the homogeneous degree in $p,q$ of the successive iterates $x_n$. A first result is that all odd values of $N$ lead to an exponential degree growth, which means that all these cases are non-integrable. The same is true for values of $|N|>4$. Thus only five values of $N$ remain, namely $N=0,\pm2,\pm4$. Next we investigate the role of the parameter $f$ and find that unless $f^2=1$ we have again an exponential growth. In the case  $N=\pm2$ the sign of $f$ is immaterial, since it can be be absorbed by a redefinition of $x$ and $z$. Moreover when $|N|=4$ the choice $f=-1$ leads again to exponential growth and thus those two cases cannot be integrable. There remain finally 6 good candidates for integrability.

{\sl Case} $N=0, f=1$

We obtain the succession of degrees 0,1,2,4,6,8,10,12,14,16,18,20, \dots, which exhibits linear growth. The corresponding mapping is therefore expected to be linearisable and, in fact, to belong to the family we dubbed linearisable mappings of the third-kind [\refdef\limit]. The moniker ``third-kind'' was introduced in order to distinguish these mappings from the already known projective ones, linearisable through a Cole-Hopf transformation, and from those that belong to the Gambier family, which can be cast into a system of two homographic mappings in cascade. The linearisation of third-kind mappings will be explained in detail in section 3.

{\sl Case} $N=0, f=-1$

We obtain the very same succession of degrees as in the case $f=1$, namely  0,1,2,4,6,8,10,12,14,16,18,20,\dots . Thus this mapping is expected to be a third-kind linearisable one  as well. Still, in the case $N=0$ the sign of $f$ is not inconsequential and we therefore expect this mapping to be different from the previous one.

{\sl Case} $N=2, f=\pm1$

In this case the degree growth saturates at just the second step, i.e. we have the succession 0,1,2,2,2,\dots. This is a clear indication that the mapping belongs to the Gambier family [\refdef\gambier]. This is indeed the case. Taking for instance $f=1$ we can rewrite the mapping as $(x_{n+1}+x_n)(x_n+x_{n-1})=(z+1/z)^2(x_n^2+(z-1/z)^2)$ which is the QRT-Gambier mapping first derived in [\refdef\capel]. 

{\sl Case} $N=-2, f=\pm1$

Here we find the succession of degrees: 0,1,2,3,6,9,12,17,22,27,34,41,48,57,\dots. The growth is clearly quadratic, which indicates that the mapping is integrable but not linearizable. Taking $f=1$ we find indeed the QRT-type invariant
$$K={(x_n+x_{n-1})(x_nx_{n-1}+z^4-z^2-1/z^2+1/z^4)\over (x_nz^2-x_{n-1})(x_n-z^2x_{n-1})+(z^4-1)^2/z^2}.\eqdef\eqix$$ 

{\sl Case} $N=4, f=1$

A cursory examination of the mapping shows that it is explicitly linear and thus trivially integrable.

{\sl Case} $N=-4, f=1$
The succession of degrees, 0,1,1,2,3,5,6,9,11,14,17,21,24,29,33,38,\dots with quadratic growth, indicates that the mapping should be integrable. We find indeed a QRT-type invariant
$$K={(x_n^2+(z^2-1/z^2)^2)(x_{n-1}^2+(z^2-1/z^2)^2)\over (x_nz^2-x_{n-1})(x_n-z^2x_{n-1})+(z^4-1)^2/z^2}.\eqdef\eqx$$ 
Thus, for the two cases $N=-2$ and $N=-4$ we expect that upon deautonomisation they will lead to discrete Painlev\'e equations. This will be the subject of section 4.

\bigskip
3. {\scap The linearisable mappings}
\medskip

We first examine the mapping corresponding to $N=0$ and $f=1$. In order to deautonomise it we consider the general form
$${(x_{n+1}z_{n+1}z_n-x_{n})(x_{n-1}z_{n-1}z_n-x_{n})-(z_{n+1}^2z_n^2-1)(z_{n-1}^2z_n^2-1)\over(x_{n+1}-z_{n+1}z_nx_{n})(x_{n-1}-z_{n-1}z_nx_{n})-(z_{n+1}^2z_n^2-1)(z_{n-1}^2z_n^2-1)/(z_{n+1}z_n^2z_{n-1})}=g_n,\eqdef\eqxi$$ 
where $g_n$ is expected to be a function of the independent variable alone. We study the degree growth of (\eqxi) and require that it be identical to the one obtained in the autonomous case. We find that in order to satisfy these constraints one has to introduce a free function $q_n$ whereupon $z_n$ and $g_n$ are given by
$$z_n=q_{n+1}q_{n-1}\quad{\rm and}\quad g_n={q_{n+2}q_{n-2}\over q_n^2}.\eqdef\eqxii$$
In order to integrate (\eqxi) we introduce the linear mapping
$$\displaylines{\big(k+(q_nq_{n-1}-1/(q_nq_{n-1}))^2\big){x_{n+1}q_{n-1}q_{n}q_{n+1}q_{n+2}-x_n\over q_{n-1}^2q_{n}^2q_{n+1}^2q_{n+2}^2-1}-\big(k-2+1/(q_nq_{n-1})^2+1/(q_nq_{n+1})^2\big)x_n\hfill\cr\hfill+\big(k+(q_nq_{n+1}-1/(q_nq_{n+1}))^2\big){x_{n-1}q_{n-2}q_{n-1}q_{n}q_{n+1}-x_n\over q_{n-2}^2q_{n-1}^2q_{n}^2q_{n+1}^2-1}=0\quad\eqdisp\eqxiii\cr}$$
and take its discrete derivative with respect to $k$ obtaining a third order mapping for $x_n$. However (\eqxi) does not possess any visible parameter which allows us to take a discrete derivative. So we obtain $x_{n-1}$ and $x_{n+2}$ from (\eqxi) and its up-shift respectively, expressed in terms of $x_{n+1},x_n$ and show that the third-order equation obtained from (\eqxiii) is identically satisfied. As we explained in [\limit] it is possible to construct the solution of (\eqxi) starting from the solution of the linear equation (\eqxiii). In order do this one starts from an initial condition where two $x$'s are given, say $x_{n-1}$ and $x_n$. Using (\eqxi) one obtains $x_{n+1}$ which allows one to compute the value of $k$ in (\eqxiii). It suffices now to solve the linear equation in order to obtain $x_n$ for all $n$ and thus construct the solution of (\eqxi).

We turn now to the case $N=0$ and $f=-1$. In the autonomous case it has the simple form
$$x_{n+1}x_{n-1}-{2\over z^2+1/z^2}x_n(x_{n+1}+x_{n-1})+x_n^2-( z^2-1/z^2)^2=0.\eqdef\eqxiv$$
At this point it is interesting to construct the invariant for (\eqxiv). It has the form
$$K=\left(x_n^2+x_{n-1}^2-4x_nx_{n-1}/(z^2+1/z^2)-( z^2-1/z^2)^2\over x_n^2+x_{n-1}^2-(z^2+1/z^2)x_nx_{n-1}+( z^2-1/z^2)^2\right)^2,\eqdef\eqxv$$
which is not bi-quadratic in $x_n$ and $x_{n-1}$. Thus the mapping (\eqxiv) is not of QRT type [\refdef\qrt],  but rather belongs to the HKY family introduced by Hirota, Kimura and Yahagi [\refdef\hky] in parallel with our findings in [\miura]. 

Before proceeding to the deautonomisation of (\eqxiv) we first present the integration of its autonomous form. We introduce the quantity
$$C_n={x_{n+1}-\lambda x_n\over x_{n-1}-\lambda x_n},\eqdef\eqxvi$$
where $\lambda=(z^2+1/z^2)/2$, 
and find that, when (\eqxiv) is satisfied, $C_n$ satisfies the equation $C_nC_{n+1}=1$. Thus $C_n=C_0^{(-1)^n}$ and the solution of (\eqxiv) can be obtained from the linear equation
$$x_{n+1}-\lambda x_n-C_0^{(-1)^n}(x_{n-1}-\lambda x_n)=0.\eqdef\eqxvii$$
The deautonomisation of (\eqxiv) proceeds along the same lines as that of (\eqxi). We start from a form (\eqxi) and by requiring that the degree growth be identical to that of the autonomous case we find the $z_n$ and $g_n$ are now given by
$$z_nz_{n-1}=q_{n+1}q_{n-1}\quad{\rm and}\quad g_n=-{q_{n+2}q_{n-1}\over q_{n+1}q_n},\eqdef\eqxviii$$
where $q_n$ is again a free function of $n$. The linearisation of this non-autonomous form cannot proceed along the same lines as that of the autonomous one, as it leads to prohibitively lengthy calculations, and we adopt a slightly different strategy. We start by remarking that introducing the auxiliary variable $y$ by $y_n=x_{n+1}/x_n$ we can transform the third-kind mapping (\eqxiv) to a Gambier type equation 
$$y_{n+1}y_{n-1}y_n(y_n-\lambda)-\lambda y_{n-1}(y_n^2-1)+\lambda y_n-1=0.\eqdef\eqxix$$
Using the same ansatz for $y_n$ we can now write the non-autonomous form of the Gambier equation as
$$\displaylines{y_{n+1}y_{n-1}y_nq_{n+3}q_{n+2}(q_{n}^2q_{n+1}^2+1)(q_{n-1}^2q_{n+1}^2-1)\big((q_{n}^2+q_{n+1}^2)q_{n+2}y_n-q_n(q_{n+2}^2q_{n+1}^2+1)\big)\hfill\cr\hfill
-y_{n-1}q_{n+1}\big(q_{n+2}^2(q_{n-1}^2q_{n+1}^2-1)(q_{n}^2q_{n+1}^2+1)(q_{n}^2q_{n+3}^2+1)y_n^2-q_{n}^2(q_{n+1}^2q_{n+3}^2-1)(q_{n+2}^2q_{n+1}^2+1)(q_{n-1}^2q_{n+2}^2+1)\big)
\hfill\cr\hfill
-y_{n-1}y_nq_nq_{n+1}q_{n+2}\big((q_{n+3}^2-q_{n-1}^2)(1+q_n^2q_{n+1}^2)(1+q_{n+1}^2q_{n+2}^2)+(q_{n+2}^2-q_{n}^2)(q_{n+1}^2q_{n-1}^2-1)(q_{n+1}^2q_{n+3}^2-1)\big)
\hfill\cr\hfill
+q_{n}q_{n-1}(q_{n+2}^2q_{n+1}^2+1)(q_{n+3}^2q_{n+1}^2-1)\big((q_{n}^2q_{n+1}^2+1)q_{n+2}y_n-q_n(q_{n+2}^2+q_{n+1}^2)\big)=0
\quad\eqdisp\eqxx\cr}$$
The way to integrate this equation is to introduce a quantity $w_n$ in the form
$$w_n={y_ny_{n-1}+a_ny_{n-1}+d_n\over f_ny_ny_{n-1}+a_ny_{n-1}+1},\eqdef\eqxxi$$
and ask that it obeys a recursion relation of the form 
$$w_{n+1}={h_nw_n+k_n\over w_n+m_n}.\eqdef\eqxxii$$
We remark that (\eqxxi) and (\eqxxii) form a Gambier system: the quantity $w_n$ obeys a homographic equation while the equation for $y_n$ is also a homographic one, the coefficients of which depend linearly on $w_n$. Requiring that the system (\eqxxi) and (\eqxxii) be equivalent to (\eqxx) leads to the following values of the coefficients
$$a_n=-{q_{n+1}(q_{n-1}^2q_{n+2}^2+1)\over q_{n-1}(q_{n+1}^2+q_{n+2}^2)},\eqdef\eqxxiii$$
$$d_n={q_{n+1}q_{n+2}(q_{n}^2+q_{n+1}^2)(q_{n-1}^2q_{n+2}^2+1)-q_{n}q_{n-1}(q_{n+1}^2+q_{n+2}^2)(q_{n+1}^2q_{n+2}^2+1)\over q_{n+2}q_{n+1}(q_{n-1}^2q_{n}^2+1)(q_{n+1}^2+q_{n+2}^2)},\eqdef\eqxxiv$$
$$f_n={q_{n+1}q_{n+2}(q_{n-1}^2q_{n+1}^2-1)(q_{n+2}^2-q_n^2)\over q_nq_{n-1}(q_{n+1}^2+q_{n+2}^2)(q_{n+1}^2q_{n+2}^2+1)},\eqdef\eqxxv$$
together with
$$h_n=d_{n+1},\eqdef\eqxxvi$$
$$k_n=-d_nd_{n+1}+(1-d_n){q_{n}q_{n+1}(q_{n+1}^2q_{n+3}^2-1)(q_{n-1}^2q_{n+2}^2+1)\over q_{n+2}q_{n+3}(q_{n-1}^2q_{n+1}^2-1)(q_{n+1}^2q_{n}^2+1)},\eqdef\eqxxvii$$
$$m_n=f_{n+1}(k_n+d_nd_{n+1})-d_n.\eqdef\eqxxviii$$
Thus equations (\eqxxi) and (\eqxxii) provide the linearisation of (\eqxx) and once $y_n$ is known $x_n$ can be constructed from the relation $x_{n+1}=y_nx_n$.

Finally we examine the Gambier mapping $N=2$ and $f=1$. As a matter of fact the deautonomisation of this mapping was already proposed in [\refdef\mickens] under the form
$${(x_{n+1}z_{n+1}z_n-x_{n})(x_{n-1}z_{n-1}z_n-x_{n})-(z_{n+1}^2z_n^2-1)(z_{n-1}^2z_n^2-1)\over(x_{n+1}-z_{n+1}z_nx_{n})(x_{n-1}-z_{n-1}z_nx_{n})-(z_{n+1}^2z_n^2-1)(z_{n-1}^2z_n^2-1)/(z_{n+1}z_n^2z_{n-1})}=z_{n+1}z_{n-1},\eqdef\eqxxix$$ 
where $z_n$ is a free function of $n$. By introducing the new variable $y_n$ through $x_n=y_n(z_n-1/z_n)$ we were able to bring (\eqxxix) to the form
$$(y_{n+1}+y_n)(y_n+y_{n-1})={(z_{n+1}^2z_n^2-1)(z_{n-1}^2z_n^2-1)\over z_n^2(z_{n+1}^2-1)(z_{n-1}^2-1)}\big(y_n^2+1\big),\eqdef\eqxxx$$
which is the general non-autonomous form of the Gambier mapping obtained in [\capel] where we have also presented its integration.

\bigskip
4. {\scap The discrete Painlev\'e equations}
\medskip

Before proceeding to the analysis of the two mappings that give rise to discrete Painlev\'e equations, we should make a general remark that is valid for both systems under consideration. Given the structure of the equations and in particular the fact that the right-hand side is an even power of $z$, a special gauge freedom exists: multiplying every $x_n$ by an arbitrary sign and $z_n$ by {\sl the same} sign, leaves the equation invariant. 

We start by analysing the mapping obtained for $N=-2$ and $f=1$. In order to deautonomise it we use the form already encountereed in the previous section namely 
$${(x_{n+1}z_{n+1}z_n-x_{n})(x_{n-1}z_{n-1}z_n-x_{n})-(z_{n+1}^2z_n^2-1)(z_{n-1}^2z_n^2-1)\over(x_{n+1}-z_{n+1}z_nx_{n})(x_{n-1}-z_{n-1}z_nx_{n})-(z_{n+1}^2z_n^2-1)(z_{n-1}^2z_n^2-1)/(z_{n+1}z_n^2z_{n-1})}=g_n,\eqdef\eqxxxi$$ 
and, assuming that $z_n$ and $g_n$ are functions of $n$, we require that the degree growth be the same as in the autonomous case. We find readily that $g_n$ must be of the form
$$g_n={1\over z_{n+1}z_{n-1}},\eqdef\eqxxxii$$
and that $z_n$ must satisfy the constraint
$$z_{n+4}z_{n-1}=\pm z_{n+2}z_{n+1}.\eqdef\eqxxxiii$$
However, given the gauge freedom we have introduced at the beginning of this section, the $\pm$ sign in (\eqxxxiii) is immaterial. It can be taken as $+$ without loss of generality.

The solution of relation (\eqxxxiii) in that case is $\log z_n=\alpha n+\beta+\phi_2(n)+\phi_3(n)$, where  $\phi_m$ is a periodic function $\phi_m(n+m)=\phi_m(n)$ with period $m$. It is given by
 $$ \phi_m(n)=\sum_{\ell=1}^{m-1} \epsilon_{\ell}^{(m)} \exp\left({2i\pi \ell n\over m}\right),\eqdef\eqxxxiv$$
where the summation starts at 1 instead of 0 and thus the constant term is absent. The latter can be absorbed through a redefinition of the secular term $\alpha n+\beta$, and thus $\phi_m$ introduces $(m-1)$ parameters. On the face of these results one could conclude that the discrete Painlev\'e equation (\eqxxxi)-(\eqxxxii) is a four parameter one. However, as we shall see, the situation is more complicated. 

In [\ancil] we investigated the various forms of additive Painlev\'e equations one can obtain starting from the generic E$_8^{(1)}$  one, by implementing limits and simplifications. The same procedure applied to multiplicative, rather than additive, equations will necessarily lead to similar results. Here we shall limit ourselves to the examination of the specific  case that is pertinent to our study of equation (\eqxxxi)-(\eqxxxii). Our starting point is the factorised form (\eqvi) obtained by the introduction of the ancillary variable $\xi_n$ through $x_n=\xi_n+1/\xi_n$. We demand that the right-hand side of (\eqvi) simplifies so that it becomes a ratio of products of two factors. At the autonomous limit we have $(\xi_n-z^3 \kappa)(\xi_n-z^3/\kappa)/((z^3\kappa\xi_n-1)(z^3\xi_n/\kappa-1))$ which, when expressed in the original dependent variable, leads to a right-hand side of the form
$$R(x_n)={zx_n-z^4(\kappa+1/\kappa)\over z^3x_n-(\kappa+1/\kappa)}.\eqdef\eqxxxv$$
We remark readily that the equation with a right-hand side equal to $1/z^2$ we used in our ansatz is obtained from (\eqxxxv) for the special value $\kappa=i$. The deautonomisation of (\eqxxxi) with right-hand (\eqxxxv) can be obtained following the method described in [\ancil].  We start from the factorised equation
$${x_{n+1}-{\xi_nz_n^{-1}z_{n+1}^{-1}}-{\xi_n^{-1}z_nz_{n+1}}\over x_{n+1}-\xi_nz_nz_{n+1}-{\xi_n^{-1}z_n^{-1}z_{n+1}^{-1}}}\,{x_{n-1}-{\xi_nz_n^{-1}z_{n-1}^{-1}}-{\xi_n^{-1}z_nz_{n-1}}\over x_{n-1}-\xi_nz_nz_{n-1}-{\xi_n^{-1}z_n^{-1}z_{n-1}^{-1}}}={(\xi_n-z_n\mu_n)(\xi_n-z_n\lambda_n)\over(z_n\mu_n\xi_n-1)(z_n\lambda_n\xi_n-1)},\eqdef\eqxxxvi$$
where $\mu_n, \lambda_n$ are related through $\mu_n\lambda_n\mu_{n+1}\lambda_{n+1}=z_{n+2}^2z_{n+1}^2z_{n}^2z_{n-1}^2$, and we use the singularity confinement approach to obtain an integrable deautonomisation. The singularity patterns we are going to follow are one where we enter the singularity through $\xi_n=z_n\mu_n$ and exit through $z_{n+3}\lambda_{n+3}\xi_{n+3}=1$ and a similar one where we permute  $\mu_n$ and $\lambda_n$. The confinement constraints are $\mu_n\lambda_{n+3}=\lambda_{n}\mu_{n+3}=z_{n+2}^2z_{n+1}^2$, leading to the very same equation (\eqxxxiii) for $z_n$. For $\mu_n, \lambda_n$ we find $\mu_n=z_{n-1}z_{n+1}\exp(\tilde\phi_3(n))\kappa$ and $\lambda_n=z_{n-1}z_{n+1}\exp(-\tilde\phi_3(n))/\kappa$. Here $\tilde\phi_3(n)$ is a periodic function, with period 3, different from the one which appears in the solution for $z_n$ and $\kappa$ is a constant. Thus the total number of parameters entering the equation at hand is 7, which suggests that this equation is associated to affine Weyl group E$_7^{(1)}$. Notice that if we take $\kappa=i$ (and put $\tilde\phi_3(n)$ to zero) the right-hand side of the equation collapses to expression (\eqxxxii). However this special value of $\kappa$ can, in principle, be modified by the application of Schlesinger transformations (which also allow to recover a nonzero $\tilde\phi_3(n)$) and thus the full freedom of the equation $N=-2, f=1$ equation is the one we just derived, involving 7 parameters. 

We turn now to the equation obtained for $N=-4$ and $f=1$, which we again deautonomise by assuming a form (\eqxxxi), while  looking for the same degree growth as in the autonomous case. This results into a right-hand side of the form 
$$g_n={1\over z_{n+1}z_n^2z_{n-1}},\eqdef\eqxxxvii$$
while $z_n$ satisfies the equation
$$z_{n+5}z_{n+4}z_{n}z_{n-1}=z_{n+3}z_{n+2}^2z_{n+1}.\eqdef\eqxxxviii$$
The characteristic equation of the latter is $k^6+k^5-k^4-2k^3-k^2+k+1=0$, the roots of which are 1(double), -1(double), $j$ and $j^2$. Thus $z_n$ has the form $\log z_n=\alpha n+\beta+\gamma n(-1)^n+\phi_2(n)+\phi_3(n)$. The term $\gamma n(-1)^n$, while atypical at first sight, does play an essential role. It suffices to remark that since only the product of $z_n$ with consecutive indices appear in the equation, $\phi_2(n)$ drops out and the contribution of the $\gamma n(-1)^n$ to a product like $z_nz_{n-1}$ is simply $\gamma (-1)^n$, i.e. an effective $\phi_2$ entering at the level of the product and introducing one parameter. Again a simple parameter counting would suggest that (\eqxxxi)-(\eqxxxvii) is a four parameter equation but, once more, things are more complicated.

We go back to the factorised from of the generic E$_8^{(1)}$-associated equation and do not take any limit as we did in the previous paragraph, but rather content ourselves with simplifications. Again we demand that the right-hand of (\eqvi) simplifies so that it becomes a ratio of products of two factors. However, at the autonomous limit, the relation between the two parameters is now such that we find the form $(\xi_n-z^4 \kappa)(\xi_n-z^4/\kappa)/((z^4\kappa\xi_n-1)(z^4\xi_n/\kappa-1))$ and an expression for $R(x_n)$:
$$R(x_n)={x_n-z^4(\kappa+1/\kappa)\over z^4x_n-(\kappa+1/\kappa)}.\eqdef\eqxxxix$$
Taking the special value $\kappa=i$ leads, as expected, to a right-hand side equal to $1/z^4$. In order to deautonomise (\eqxxxi) with right-hand side (\eqxxxix) we start from the same factorised form (\eqxxxvi) where now the $\mu_n, \lambda_n$ obey the relation $\mu_n\lambda_n=z_{n+1}^2z_{n}^2z_{n-1}^2$. We shall not enter into the details of the singularity confinement analysis but immediately give the confinement constraints: $\mu_n\lambda_{n+4}=\mu_{n+4}\lambda_{n}=z_{n+1}^2z_{n+2}^2z_{n+3}^2$. Finally, the solution for the $\mu_n, \lambda_n$ is $\mu_n=z_{n-1}z_nz_{n+1}\exp(\phi_4(n))\kappa$ and $\lambda_n=z_{n-1}z_nz_{n+1}\exp(-\phi_4(n))/\kappa$. Notice that, since only the products $z_n\mu_n$ and $z_n\lambda_n$ appear in the equation, the effect of the $\gamma (-1)^n$ term in $z_n$ is again that of an effective $\phi_2$ introducing one parameter. Thus the full equation obtained for $N=-4, f=1$ has 8 parameters and is associated to the affine Weyl group E$_8^{(1)}$ [\ohta], equation (\eqxxxi)-(\eqxxxvii) being just a special case thereof.

\bigskip
5. {\scap Conclusions}
\medskip

In this paper we set out to derive discrete integrable equations of multiplicative type, the form of which is inspired by multiplicative discrete Painlev\'e equations associated with the affine Weyl group  E$_8^{(1)}$. In this sense the paper is a sequel to our recent paper which dealt with equations of additive type  [\miura]. Our investigation here, based on the algebraic entropy method, yielded 6 integrability candidates, four linearisable ones (among which one that is trivially linear) and  two which exhibit quadratic degree growth and,  upon deautonomisation, were therefore expected to lead to discrete Painlev\'e equations. 

All three nontrivial linearisable mappings were deautonomised and integrated. The interesting result here was the case of the $N=0, f=-1$ mapping which is not of QRT type but rather belongs to the family introduced by Hirota, Kimura and Yahagi (HKY). The integration of its nonautonomous form was somewhat tricky and the only way we found to perform this integration was to go from the third-kind mapping to the associated Gambier one and integrate the latter. For the two mappings which were expected to lead to a discrete Painlev\'e equation, the deautonomisation under the initial simplified form led to systems that are rather poor in their number of free parameters. This spurred a careful examination of the systems at hand which showed that the multiplicative ansatz we had been working with was artificially constrained. Namely, the form we were working with was a special case of a richer one in which the values of some parameters had been fixed. Going back to the more general form and performing its deautonomisation we were able to show that the two discrete Painlev\'e equations obtained possessed 7 and 8 parameters, respectively, and therefore should be associated to the affine Weyl groups E$_7^{(1)}$ and E$_8^{(1)}$.

A final cautionary remark is in order here. Usually when working with generic equations associated to some affine Weyl group which can support both additive and multiplicative systems, it suffices to obtain the results in the simpler case, usually the additive one, and then transcribe them to the other case. This works in the case of the E$_8^{(1)}$ group for elliptic equations as well. However, when one works with ad hoc simplified cases, as in the case of our previous paper and the present one, this correspondence between the additive and multiplicative families clearly breaks down and one must perform the analysis afresh. 

\vfill\eject
\bigskip
{\scap Acknowledgements}
\medskip

RW would like to acknowledge support from the Japan Society for the Promotion of Science (JSPS), through the JSPS grant: KAKENHI grant number 15K04893. 

\bigskip
{\scap References}
\medskip

 [\dps] A. Ramani, B. Grammaticos and J. Hietarinta, Phys. Rev. Lett. 67 (1991) 1829.
  
 [\desoto] B. Grammaticos, F.W. Nijhoff and  A. Ramani, {\it Discrete Painlev\'e equations}, in The Painlev\'e property -- One Century later, R. Conte (Ed.), New York: Springer-Verlag, (1999) p. 413.
 
 [\sincon] B. Grammaticos, A. Ramani and V. Papageorgiou, Phys. Rev. Lett. 67 (1991) 1825.

 [\hiv] J. Hietarinta and C-M. Viallet, Phys. Rev. Lett. 81, (1998) 325.
 
 [\algeo] T. Mase, R. Willox, B. Grammaticos and   A. Ramani, Proc. Roy. Soc. A 471 (2015) 20140956.
 
 [\sakai] H. Sakai, Commun. Math. Phys. 220 (2001) 165.

 [\grand] Y. Ohta, A. Ramani and B. Grammaticos, J. Phys. A 34 (2001) 10523.
 
 [\ohta] Y. Ohta, A. Ramani and B. Grammaticos, J. Phys. A 35 (2002) L653.

 [\murata] M. Murata, Funkcialaj Ekvacioj 47 (2004) 291.
 
 [\yoneda] M. Murata, H. Sakai and J. Yoneda, J. Math. Phys. 44 (2003) 1396.

 [\tsuji] M. Noumi, S. Tsujimoto and Y. Yamada, {\sl Pad\'e interpolation for elliptic Painlev\'e equation}, in Symmetries, integrable systems and representations, K. Iohara et al. (eds.), Springer Proc. Math. Stat., 40 (2013) 463.

 [\addeight] A. Ramani and B. Grammaticos, J. Phys. A 48 (2015) 355204.
 
 [\ellip] B. Grammaticos and A. Ramani, J. Phys. A 48 (2015) 16FT02.
 
 [\trijmp] B. Grammaticos and A. Ramani, J. Math. Phys. 56 (2015) 083507.
 
 [\ancil] A. Ramani and B. Grammaticos, J. Phys. A 50 (2017) 055204.

 [\ellip] B. Grammaticos and A. Ramani, J. Phys. A 49 (2016) 45LT02.
 
 [\kanoya] K. Kajiwara, M. Noumi and Y. Yamada, J. Phys. A 50 (2017) 073001.
 
 [\miura] A. Ramani, B. Grammaticos and R. Willox, J. Math. Phys. 58  (2017) 043502.
 
 [\limit] A. Ramani, B. Grammaticos and Y. Ohta,  Nonlinearity 13 (2000) 1073.
 
 [\gambier] B. Grammaticos, A. Ramania and S. Lafortune Physica A 253 (1998) 260.

 [\capel] A. Ramani and B. Grammaticos, Physica A 228 (1996) 160.
 
 [\qrt] G.R.W. Quispel, J.A.G. Roberts and C.J. Thompson, Physica D34 (1989) 183.
 
 [\hky] R. Hirota, K. Kimura and H. Yahagi, J. Phys. A. 34 (2001) 10377.
 
 [\mickens] B. Grammaticos, A. Ramani, J. Satsuma and R. Willox, J. Math. Phys. 53 (2012) 023506.

\end{document}